\documentclass[a4paper]{article}

\usepackage[english]{babel}
\usepackage[utf8x]{inputenc}
\usepackage[T1]{fontenc}

\usepackage[a4paper,top=3cm,bottom=2cm,left=3cm,right=3cm,marginparwidth=1.75cm]{geometry}

\usepackage{amsmath}
\usepackage{graphicx}
\usepackage[colorinlistoftodos]{todonotes}
\usepackage[colorlinks=true, allcolors=blue]{hyperref}
\usepackage{natbib}
\usepackage{rotating}
\usepackage{authblk}

\title{Was Doggerland catastrophically flooded by the Mesolithic Storegga tsunami?}
\author[*]{Jon Hill}
\author[**]{Alexandros Avdis}
\author[**]{Simon Mouradian}
\author[**]{Gareth Collins}
\author[**,$\dag$]{Matthew Piggott}
\affil[*]{Environment Department, University of York, UK. \texttt{jon.hill@york.ac.uk}}
\affil[**]{Earth Science and Engineering, Imperial College London, UK.}
\affil[$\dag$]{Grantham Institute for Climate Change, Imperial College London, UK.}

\begin{document}
\maketitle

\begin{abstract}
Myths and legends across the world contain many stories of deluges and floods. Some of these have been attributed to tsunami events. Doggerland in the southern North Sea is a submerged landscape and is thought to have been heavily affected by a tsunami such that it was abandoned by Mesolithic human populations at the time of the event. The tsunami was generated by the Storegga submarine landslide off the Norwegian coast which failed around 8150 years ago. At this time there were also rapid changes in sea level associated with the deglaciation of the Laurentide ice sheet and drainage of its large proglacial lakes, with the largest sea level jumps occurring just prior to the Storegga event. The tsunami affected a large area of the North Atlantic and left sedimentary deposits across the region, from Greenland, through the Faroes, the UK, Norway and Denmark. From these sediments, run-up heights of up to 20 metres have been estimated in the Shetland Isles and several metres on mainland Scotland. However, sediments are not preserved everywhere and so reconstructing how the tsunami propagated across the North Atlantic before inundating the landscape must be performed using numerical models. These models can also be used to recreate the tsunami interactions with now submerged landscapes, such as Doggerland. Here, the Storegga submarine slide is simulated, generating a tsunami which is then propagated across the North Atlantic and used to reconstruct the inundation on the Shetlands, Moray Firth and Doggerland. The uncertainty in reconstructing palaeobathymetry and the Storegga slide itself results in lower inundation levels than the sediment deposits suggest. Despite these uncertainties, these results suggest Doggerland was not as severely affected as previous studies implied. It is suggested therefore that the abandonment of Doggerland was primarily caused by rapid sea level rise prior to the tsunami event.
\end{abstract}

\section{Introduction}

Global myths and \lq traditional oral tales\rq\ contain many stories of great deluges or flooding events. A number of these have been attributed to tsunami inundation \citep{Bryant2007-fm,Heaton1985-du,Antonopoulos1992-pv} or sea level rise \citep{Nunn2014-pq,Giosan2009-em} and an increasing body of geological evidence is being discovered to test the validity of the legends \citep{Bruins2008-dk}. In the southern North Sea, the now submerged island of Doggerland was still emergent at around 8 kyr and was occupied by Mesolithic communities \citep{Ballin2017-wp}. Due to rising sea levels Doggerland was eventually flooded completely \citep{Coles1998-ii}. However, studies have hypothesised that Doggerland was inundated by a large tsunami caused by the Storegga submarine landslide which occurred around 8150 years ago \citep{Bondevik2012-uy}, with some estimates of 5m high waves impacting the island \citep{Hill2014-jo}. If true, this would have been a catastrophic event, affecting the communities living there, and possibly causing the abandonment of the island \citep{Weninger2008-zc}. However, due to uncertainties in the relative timing of sea level rise events \citep{Lawrence2016-bm} and the magnitude of the Storegga tsunami on Doggerland \citep{Hill2014-jo}, it has been challenging to untangle the effect of these events on Doggerland.

\begin{figure}[ht]
\centering
\includegraphics[width=0.9\textwidth]{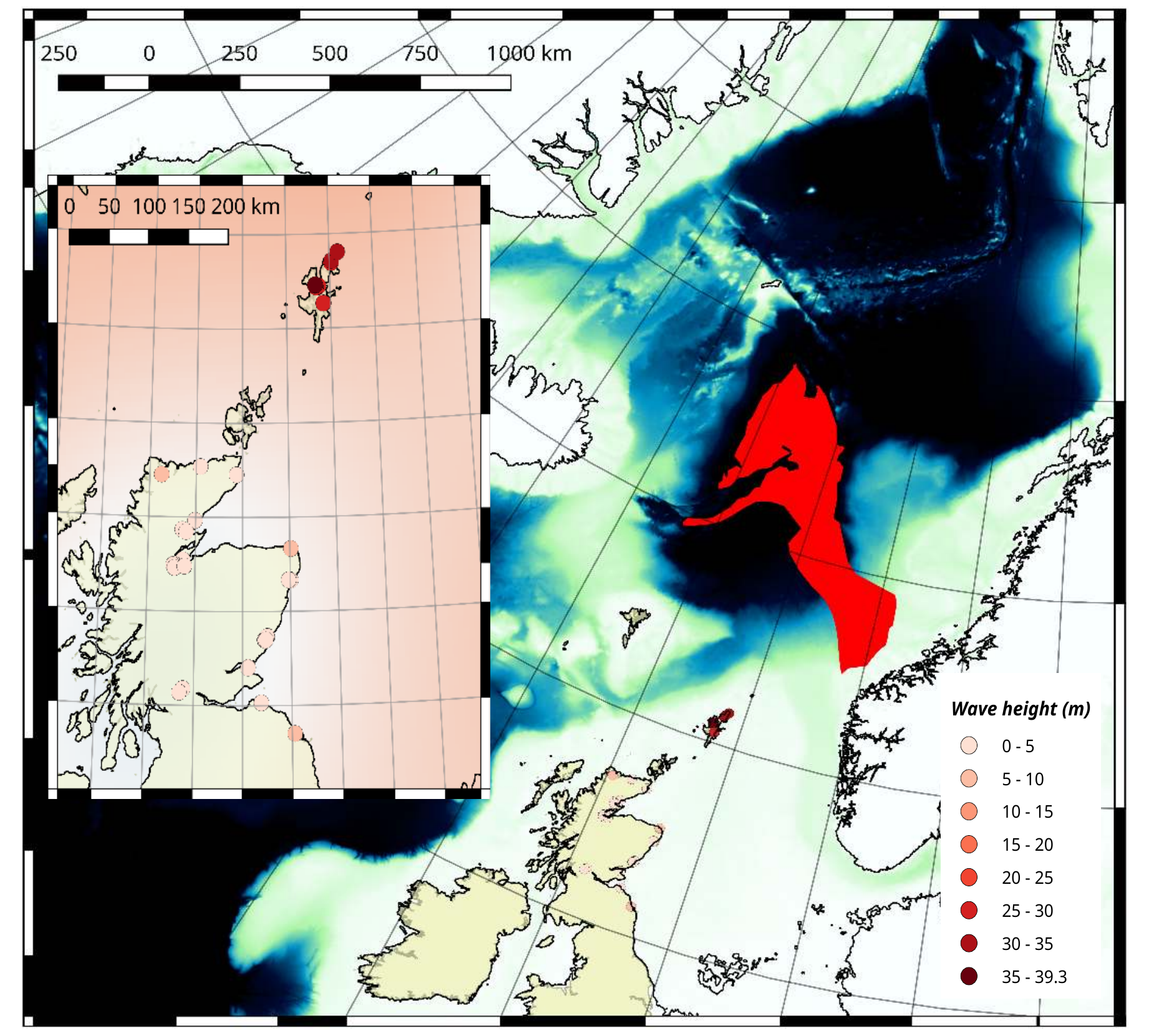}
\caption{\label{fig:overviewmap}The Storegga final deposit (red) initiated off the coast of Norway and spread into the Norwegian-Greenland sea (main map). Sedimentary deposits in the UK are found in the Shetlands and northern Scotland and as far south as northern England (circles -- main map and inset). The coastline at the time (black line) was different to today's (filled landmass -- UK only) due to changes in relative sea level.}
\end{figure}

\begin{figure}[ht]
\centering
\includegraphics[width=0.7\textwidth]{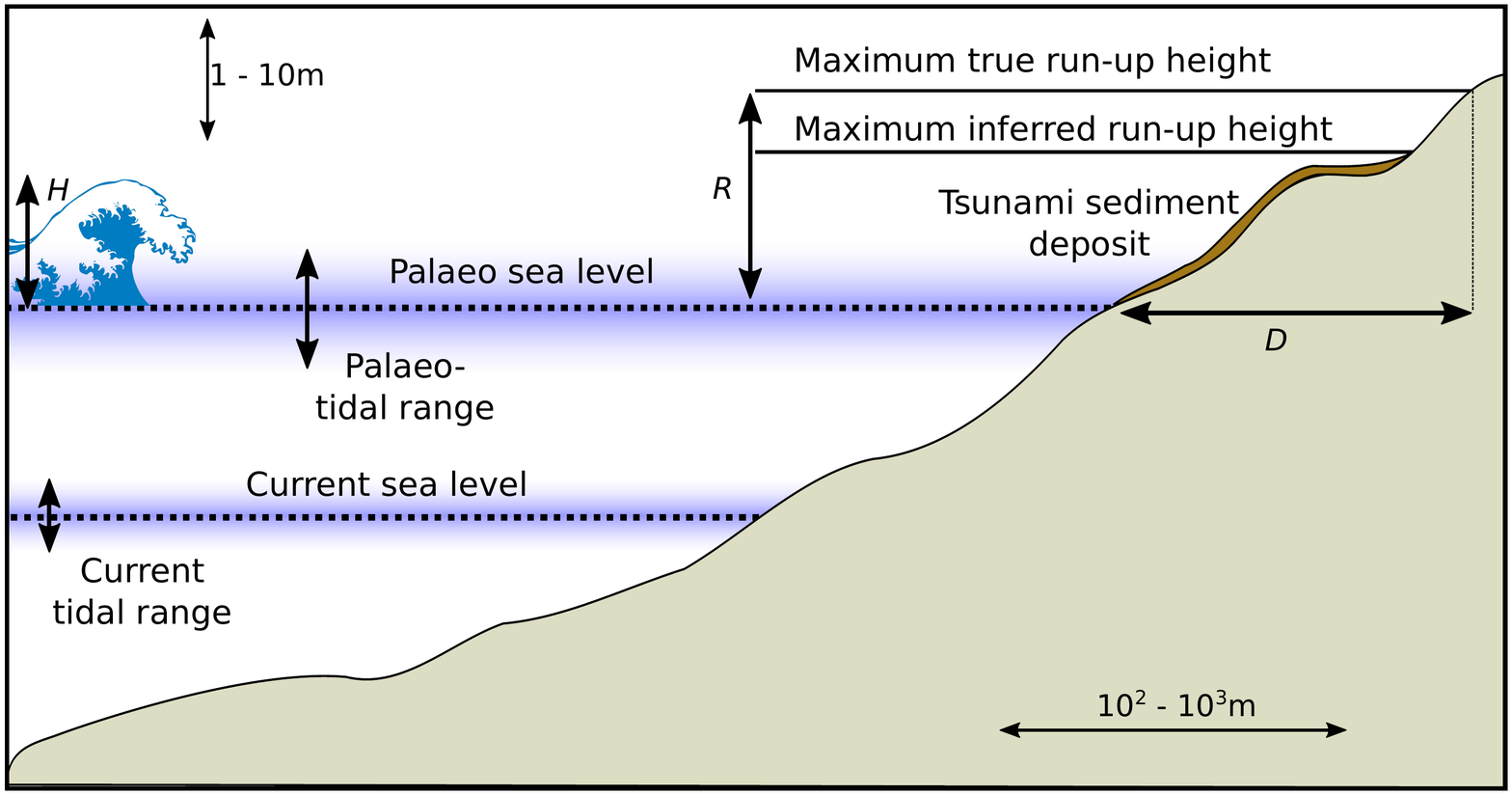}
    \caption{\label{fig:terms}Terms used to describe the magnitude of a tsunami wave when inundating land. \textit{H} is the tsunami wave height offshore, \textit{R} is the run-up height, and \textit{D} is the inundation distance. The relationship of these terms to the metrics inferred from tsunami sediment deposits is shown, as is the relationship between hypothetical current and palaeo-sea levels and tidal ranges.}
\end{figure}

The Storegga tsunami was caused by a large, 2400--3200 km$^3$ submarine landslide \citep{Haflidason2005-wz} off the Norwegian coast (Fig. \ref{fig:overviewmap}). The resulting wave spread across the Norwegian-Greenland sea, inundating the coastlines of Norway \citep{Bondevik2003-if}, the Faroe Islands  \citep{Grauert2001-sc}, Greenland \citep{Wagner2007-jt}, the UK \citep{Smith2004-xa}, and Denmark \citep{Fruergaard2015-bb}. Wave heights of tens of metres are estimated to have inundated the Norwegian coastline and nearly 20m on the Shetland Isles \citep{Bondevik2003-if}. When a tsunami interacts with the coastline, the wave steepens and then travels inland. Depending on local geomorphological characteristics, a wave can travel several kilometres inland. Typically, a tsunami is formed of several waves forming a wave-train. Each wave in this train will inundate inland depending on the size of the wave. A number of variables can be used to measure the impact of a wave (Fig. \ref{fig:terms}). Tsunamis that have been observed are typically characterised by inundation distance or run-up height. Inundation distance can be derived using eye-witness accounts, debris lines or mud deposits from the wave. Run-up height is derived from watermarks left on buildings. Wave height is typically the height of the wave just offshore, before interaction of the tsunami with the seabed starts to become the dominant control on wave height. Both run-up and wave height are taken from mean local sea level. Once the wave has inundated it can take hours or days before the water recedes \citep{Chini2012-my}. This means with a typical period of minutes to hours a tsunami wave train can inundate land that is already flooded from the previous waves. 

The estimated wave heights or run-up heights of palaeotsunamis can be derived from sedimentary deposits left by the tsunami (e.g. \citet{Smith2004-xa}). These deposits are often embedded within peat or other sedimentary layers and can be traced inland \citep{Bondevik2003-if}. Other deposits are found within lake sediments \citep{Romundset2011-qq}. From each deposit a minimum run-up can be inferred by calculating the palaeoaltitude of the site (Fig. \ref{fig:terms}) taking into account relative sea level changes from modern bathymetry/topography. The sediment deposit indicates the minimum inundation distance (and hence run-up height) as the grain size of the sediment decreases with increased distance from the shoreline as does the amount of sediment deposited. In the Japanese 2011 tsunami a 0.5 cm or greater sand layer was deposited for 57\% to 100\% of the inundation distance, depending on total inundation distance \citep{Abe2012-rw}. The deposits formed by the Storegga tsunami in the UK are mostly found in the north-east of Scotland, including the Shetland Isles (Fig. \ref{fig:overviewmap}). The Storegga deposits are generally a medium to fine sand layer, often pale in colour \citep[e.g.][]{Smith1999-wd,Dawson2000-kf}, but coarse sand to gravel in some locations \citep{Long2016-nj}, found within peat layers \citep{Smith2004-xa}. The deposits often contain peat rip-up clasts embedded within the sand deposit, indicating erosion \citep{Dawson2000-kf}. The sediment found in the tsunami deposits is clearly dependent on the local sediment characteristics, but also on the wave and local geomorphological characteristics \citep{Takashimizu2012-kr}. 

The Storegga tsunami occurred at a time of rapid sea level rise \citep{Clarke2004-eb} -- an estimated 1.4m over 500 years, which may have occurred as a series of pulses of decimetre to metre-scale \citep{Lawrence2016-bm}. The primary cause of these sea level changes was the 8.2ka cold event whereby the rapid drainage of the Laurentide proglacial lakes caused disruption to the Atlantic Meridional Overturning circulation \citep{Barber1999-ai}. These lakes were fed by the retreat of the Laurentide ice sheet, the largest of the northern hemisphere ice sheets, which in itself had a substantial effect of post-Glacial Maximum sea levels \citep{Carlson2007-mi}. There is significant debate on exactly how and when the lakes drained; whether in a single event or as multiple events, and the precise timing of the events \citep{Hijma2010-xe,Lawrence2016-bm}. However, the 8.2ka event is generally thought to have been between 8740--8160 years ago \citep{Lawrence2016-bm}, the most recent of which overlaps with the possible dates of the Storegga tsunami. These sea level variations had a major effect on Mesolithic communities including a period of migration \citep{Turney2007-nw}, which was undoubtedly related to rapid transgression in the region, with coastline moving landward over 30km at this time \citep{Hijma2011-qq}.

These rapid changes in sea level rise propagate uncertainties into the estimating of tsunami wave height from sedimentary deposits, as do uncertainties in the Glacial Isostatic Adjustment (GIA) models used, which can be of the order of metres at any given time slice \citep{Kuchar2012-lz}. In addition, when estimating the run-up height from sediment deposits several assumptions are made, including that there is no change in tidal range at the location (e.g. \citet{Smith2004-xa}) which is generally not the case \citep{Shennan2002-re}. Detailed knowledge of sedimentary changes that have occurred in the region over the last 8,000 years are lacking leading to further uncertainties, which include the preservation of coastal sediments \citep{Forbes1995-xr}.

Although sedimentary deposits can be used to infer the tsunami wave height at select locations, they cannot be used to infer waves on now submerged landscapes. Numerical models can, however. Several modelling studies of the Storegga slide and tsunami have been carried out, which varied in resolution and numerical discretisation used, but all using very similar methods to generate the initial wave \citep{Harbitz1992-eh,Bondevik2005-fe,Hill2014-jo}. All models used a solid-block wave generator whereby a fixed shape is moved at a prescribed speed downslope. Deformation in the slide which can alter the initial waveform \citep{Smith2016-dl} is therefore not considered in such models. The two earliest models used modern bathymetry and differed in the shape of the wave generator \citep{Harbitz1992-eh,Bondevik2005-fe}. The most recent modelling study includes the effect of relative sea level changes over the last 8,000 years to reconstruct the palaleobathymetry at the time improving the agreement between the modelled wave and the minimum wave heights derived from sedimentary deposits \citep{Hill2014-jo}. The model of \citet{Hill2014-jo} also used multi-scale resolution that varied spatially such that higher resolution was used along complex coastlines. However, several important effects, such as inundation, have not been considered in any of the models to date. All models had \lq cliff\rq  coastlines where the boundary conditions used did not permit flow onto the shoreline. Thus flooding from waves approaching the shores is not modelled and the wave is reflected directly from the coastline back into the domain. For this reason all previous models compared the wave height some distance from this cliff-coastline with the run-up heights derived from sediments \citep{Harbitz1992-eh,Bondevik2005-fe,Hill2014-jo}. Assuming the slope and characteristics of the bathymetry are known, it is also possible to use an empirical relationship, such as that by \citet{Synolakis2009-tg}, to derive estimated run-up heights from these. However, \citet{Long2016-nj} used such relationships for the most westerly Storegga deposit found in the UK and found the estimated run-up height was lower (4.6 m) than the run-up height estimated from sediments (8-10 m depending on the relative sea level reconstruction used).

Here, the multiscale modelling approach of \citet{Hill2014-jo} is used to simulate the generation of the Storegga slide tsunami and propagate it towards the UK, including Doggerland. That simulated wave is then used to force the boundary of a second model, with increased spatial resolution, that is capable of modelling coastal inundation. Three separate inundation models are carried out on the Shetlands, the Moray Firth and Doggerland. The first two models are used to evaluate the agreement bewtten the coupled model to minimum run-up heights derived from sediment deposits. These coupled models constitute the first attempt to validate a full reconstruction of the Storegga tsunami from initiation to final inundation against inferred runup heights from sediment deposits. The reconstruction is then be used to recreate the wave that impacted Doggerland for which there are no sedimentary records. 

\section{Methods}

\subsection{Regional-scale model}

The regional-scale tsunami model used is that of \citet{Hill2014-jo}. Briefly, this model solves the three-dimensional non-hydrostatic Navier-Stokes equations on a spherical shell under a Bousineseq assumption in a rotating reference frame \citep{Piggott2008-hq}. The equations are solved in a finite-element framework using a mixed discretisation of linear discontinuous Galerkin for velocity and a second-order continuous Galerkin formulation for pressure \citep{Cotter2009-de}. A combined pressure-free-surface boundary condition is applied on the upper surface \citep{Funke2011-cc} with coastlines and the seabed using a no-normal flow boundary condition with a velocity-dependent drag term. The simulation is a carried out on a spherical shell which is one element thick in the vertical.

The tsunami is initiated using a dynamic solid-block model that has a prescribed shape and velocity-time profile. As the slide moves water is forced upwards at the leading edge and pulled downwards at the trailing edge, creating the wave. Shear along the top of the solid-block is also included. The tsunami is then propagated over a reconstructed palaeobathymetry derived from GEBCO modern bathymetry adjusted according to the GIA model of \citet{Bradley2011-kn}. For details of the set-up and parameters used, see \citet{Hill2014-jo}.

\subsection{Local-scale inundation model}

\begin{figure}[ht]
\centering
\includegraphics[width=0.9\textwidth]{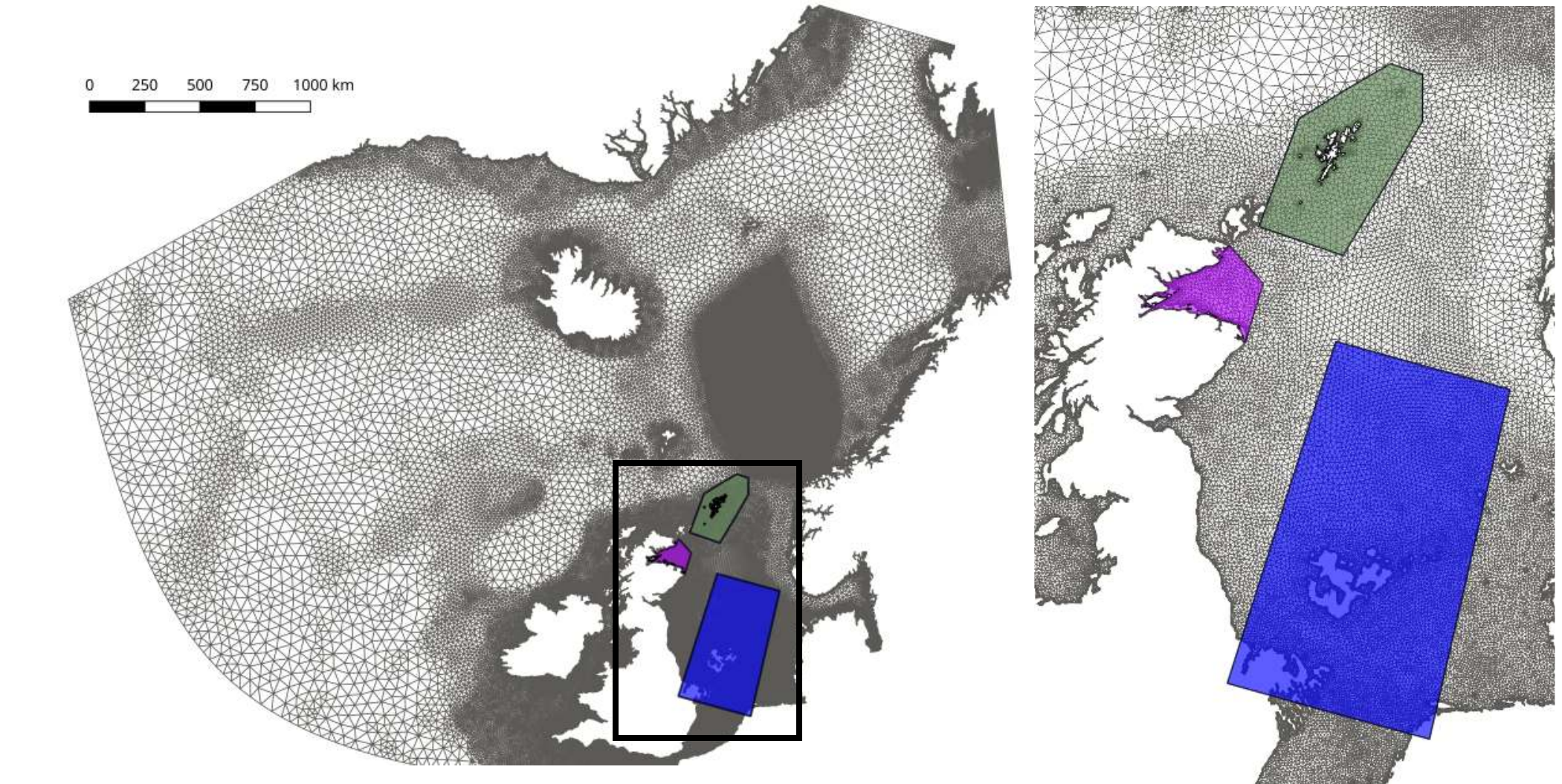}
\caption{\label{fig:domains}The regional Storegga simulation mesh, with the three local inundation sub-domains highlighted.}
\end{figure}

To model the impacts of inundation, three local-scale models were constructed for the Shetland Isles, the Moray Firth and Doggerland, respectively (Fig. \ref{fig:domains}). The wave inundation model used here is a shallow water equation model using a finite element discretisation (Telemac2D). Telemac2D has been used to simulate inundation from tsunami waves previously \citep{Horsburgh2008-bd} and has a wetting-drying scheme that performs well at high resolution \citep{Eric_Jones2006-pa}. It is capable of simulating the inundating flood wave celerity to a 3\% accuracy in dam-break simulations \citep{Hervouet2000-tl}, and has minimal numerical dispersion \citep{Malcherek2000-do}. For each model a triangular, multi-scale mesh was constructed, based on modern bathymetric/topographic data which was then adjusted using the GIA model of \citet{Bradley2011-kn} as per the regional-scale model. The mesh resolution is a simple function of water depth to roughly maintain a constant number of elements per wavelength and ensure the model complies with the CFL criterion. For each model, the boundaries of the computational domains and the mesh resolution distribution were constructed in QGIS \citep{QGIS_software} and \texttt{qmesh} \citep{qmesh:2017}. \texttt{qmesh} was then used to convert both boundary geometry and mesh size metric to a suitable UTM projection and to construct a mesh, using GMSH \citep{Geuzaine2009-dd}. Similar to \citet{Horsburgh2008-bd}, a bed friction was represented as a Nikuradse roughness length of 0.01 m for low drag simulation, 0.05 for medium drag and 0.1 for high drag. These values are representative of clean gravel, light vegetation and heavy vegetation respectively \citep{Hervouet2000-xe}. A forcing boundary condition of each of these models was based on the free surface height and depth-averaged fluid velocity from the regional-scale model. Other boundaries were either solid or free-radiation (see below for details). For each model three different palaeobathymetric reconstructions were made, each based on the modern bathymetry but adjusted using \citet{Bradley2011-kn} as in the regional model scenario plus an additional -5m, 0m or +5m sea level change applied. These changes to the sea level reconstruction account for uncertainties in the GIA reconstructions. These were run for each of the three drag scenarios giving nine model runs in total for each of the three locations. Each model was carried out in a UTM projection. Using the available sedimentary data from the Shetlands and Moray Firth regions, the model configuration that provides an inundation that most closely matches estimated wave heights can be derived and then used to assess if the Storegga tsunami was a catastrophic event on Doggerland.

-
\subsubsection{Shetland Isles model}

\begin{figure}[ht]
\centering
\includegraphics[width=0.5\textwidth]{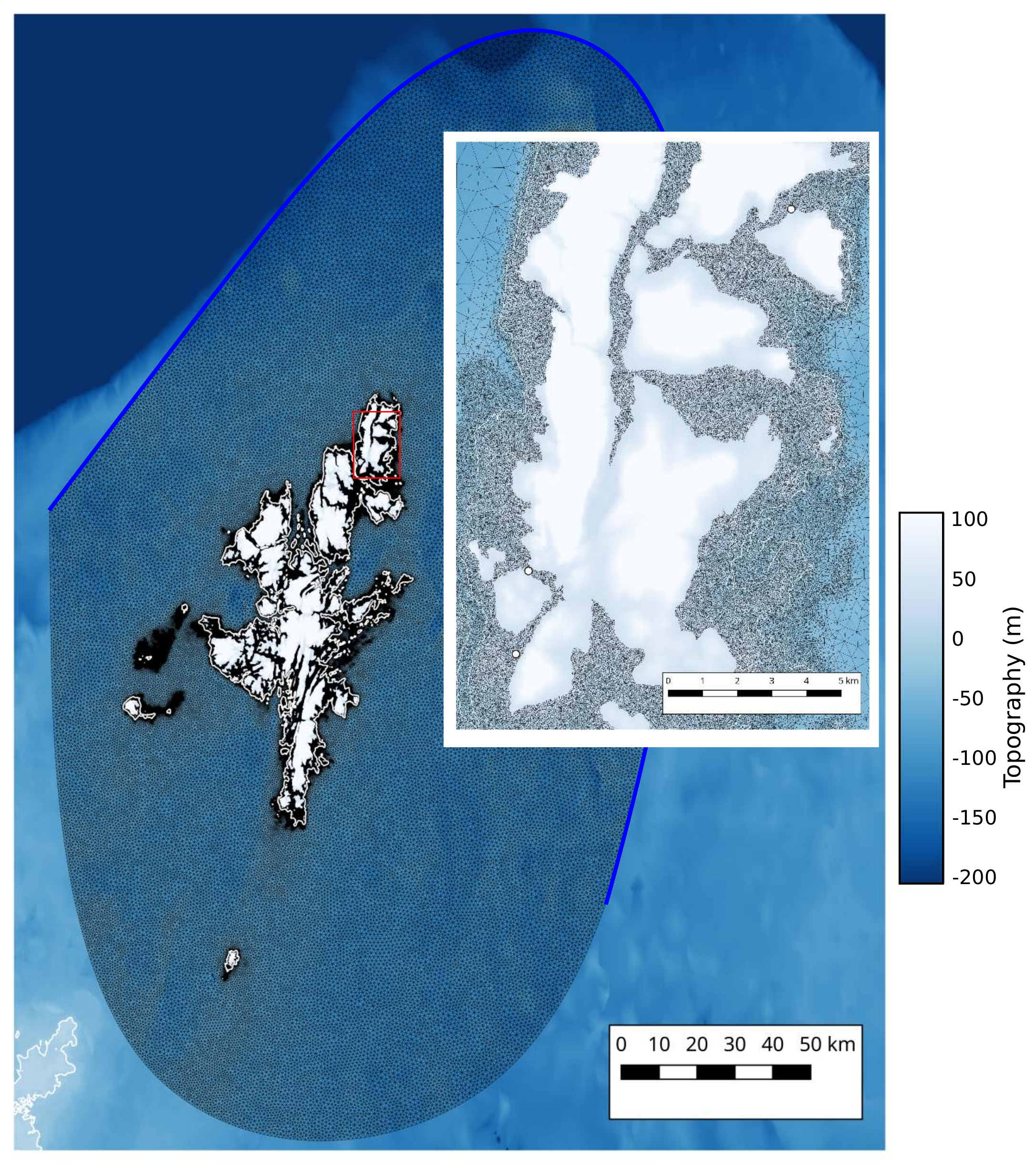}
    \caption{\label{fig:shetland_mesh}Mesh and domain used in the local tsunami inundation simulations in the Shetlands. Inset shows close up and Yell, Unst and Fetlar. Thin white line is the palaeocoastline of the simulation with no additional sea level adjustments made. Note mesh extends inland depending on the topography in the region.  Bathymetry data is \textcopyright Crown Copyright/SeaZone Solutions. All Rights Reserved. Licence No. 052006.001 31st July 2011. Not to be used for Navigation. Topographic data is  \textcopyright Crown Copyright and Database Right (2017). Ordnance Survey (Digimap Licence).}
\end{figure}

The inundation model for the Shetland Isles consisted of 30m resolution bathymetric data from SeaZone, complemented by 5m resolution topographic data from the Ordnance Survey, which was resampled to 30m resolution. Boundaries for this model were the 50m contour to reduce the modelling of topography that was too high for the wave to inundate. Other boundaries are as shown in Figure \ref{fig:shetland_mesh}. Mesh resolution varied from 60 m to 20 km, with all elements at bathymetry greater than 20 m having 60 m resolution. Bathymetry was used to alter mesh resolution according to the wave celerity ($\sqrt{gh}$) to ensure the wave could be adequately captured in shallow water. A total of 1,297,938 elements and 648,868 nodes constituted the mesh (Fig \ref{fig:shetland_mesh}).

\subsubsection{Moray Firth model}

\begin{figure}[ht]
\centering
\includegraphics[width=0.5\textwidth]{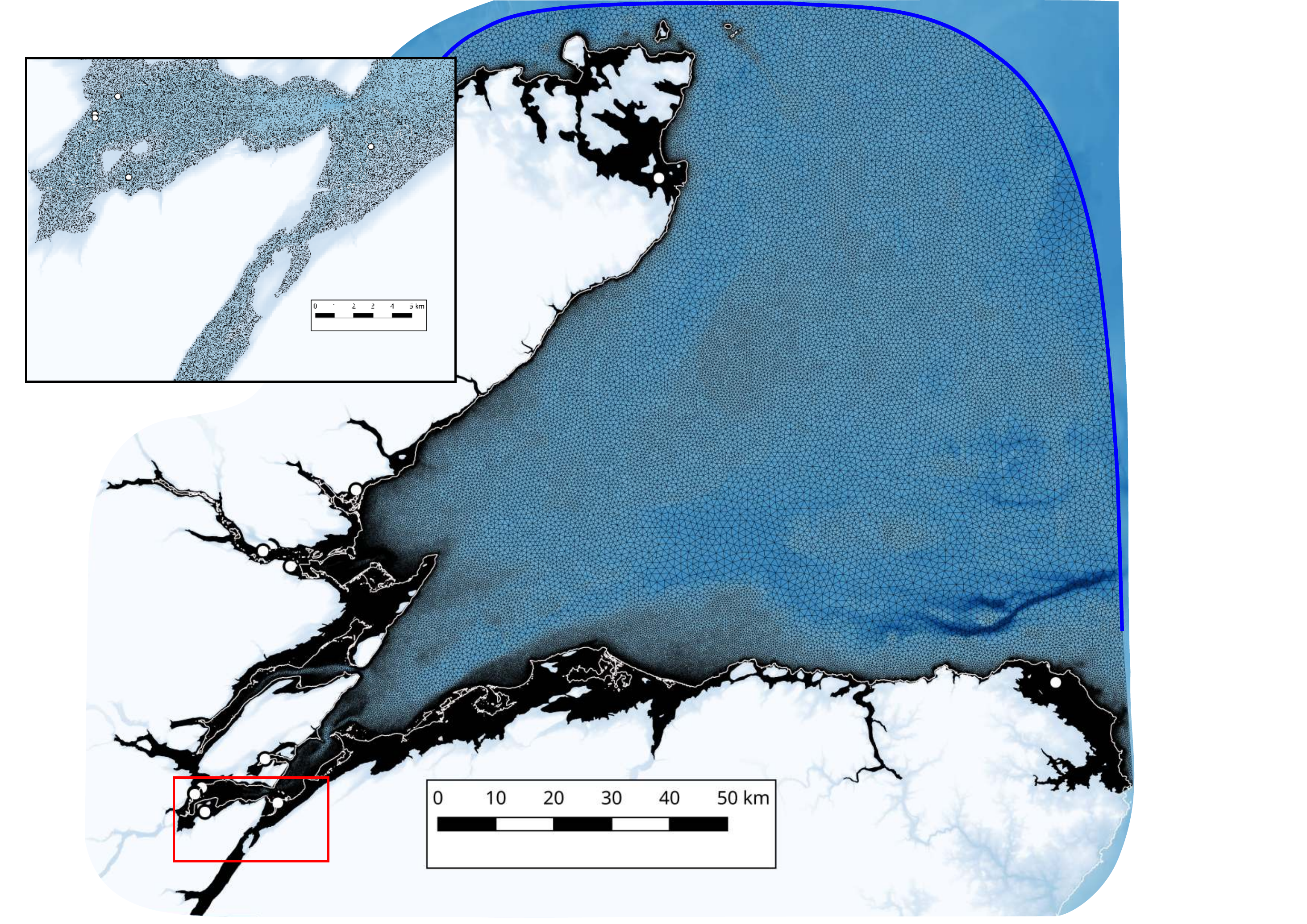}
    \caption{\label{fig:moray_mesh}Mesh and domain used in the local tsunami inundation simulations in the Moray Firth. Inset shows close up of the Inner Moray Firth and Cromarty Firth. Thin white line is the palaeocoastline of the simulation with no additional sea level adjustments made. Note mesh extends inland depending on the topography in the region. Bathymetry data is \textcopyright Crown Copyright/SeaZone Solutions. All Rights Reserved. Licence No. 052006.001 31st July 2011. Not to be used for Navigation. Topographic data is  \textcopyright Crown Copyright and Database Right (2017). Ordnance Survey (Digimap Licence).}
\end{figure}

As with the Shetland Isles model, the Moray Firth local-scale model was constructed from bathymetric data from SeaZone and topographic data from the Ordnance Survey with an approximately 30m resolution. Boundaries for this model consisted of the 40 m topographic contour and a smooth arc, from the north-western edge to the south-eastern edge, which acted as the forcing boundary for this domain. The minimum resolution used here was 45 m with a maximum resolution of 5 km. Due to the fact that the majority of the domain was shallow water, no alteration of resolution based on bathymetry was carried out. Instead resolution varied smoothly from 45 m at the 0 m topographic contour to 5 km over a 10 km distance. All areas where the water depth was 3 m or less were set to the minimum resolution and the external boundary had increased resolution at 1km which smoothly increased to 5km at 100 km distance. The mesh was constructed in a UTM 30N projection space and had 503,696 nodes and 1,007,464 elements (Fig \ref{fig:moray_mesh}). 

\subsubsection{Doggerland}

\begin{figure}[ht]
\centering
\includegraphics[width=0.5\textwidth]{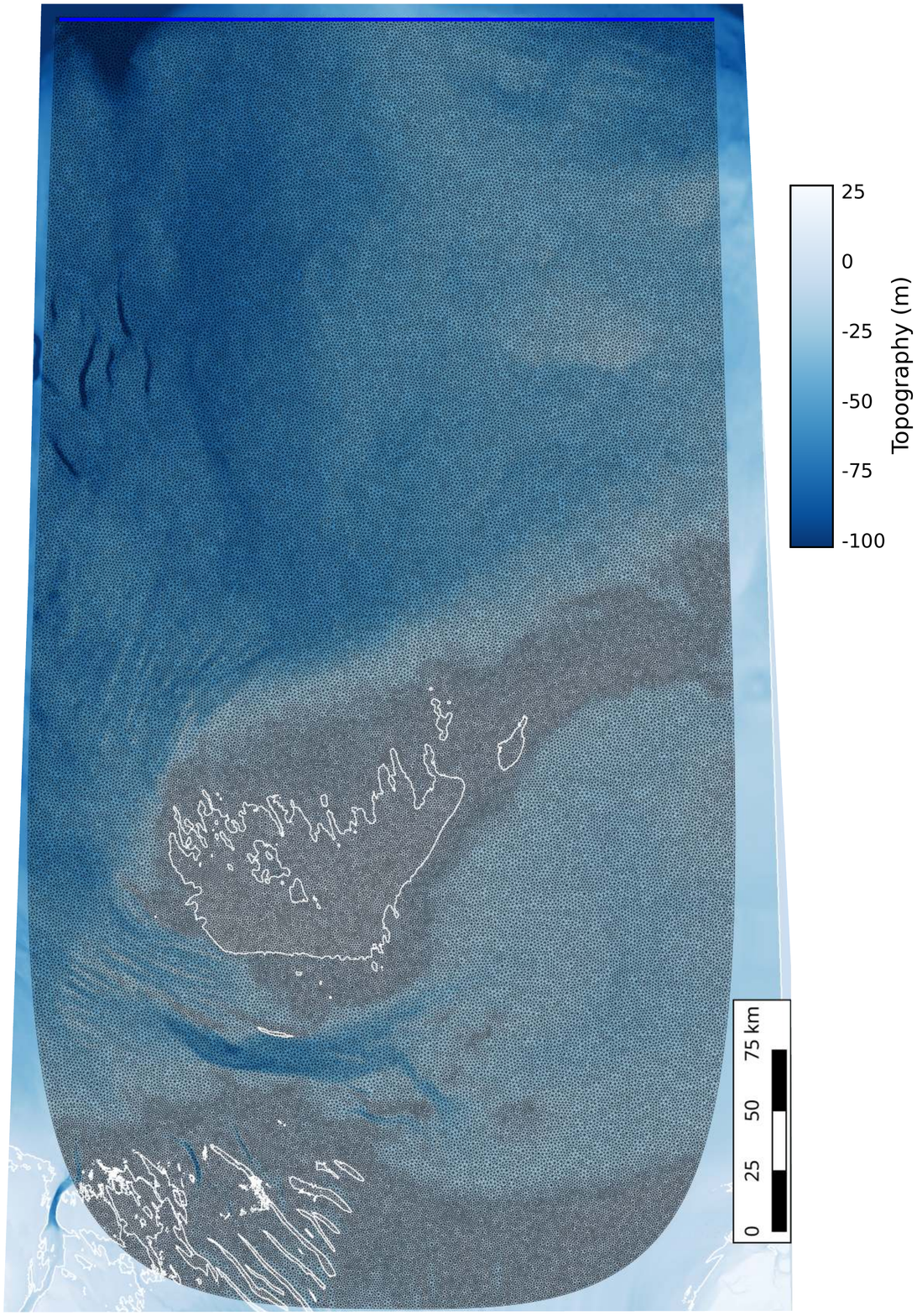}
    \caption{\label{fig:doggerland_mesh}Mesh and domain used in the local tsunami inundation simulations of Doggerland. Thin white line is the palaeocoastline of the simulation with no additional sea level adjustments made. Bathymetry is derived from EMODnet data \citep{Service2016-bq}}
\end{figure}

The area around Doggerland is on the boundary of the SeaZone data availability and the quality of data in the SeaZone dataset here is low. The bathymetry here was reconstructed from EMODnet data \citep{Service2016-bq}. Nominal resolution of the data is around 188 m. The mesh is a simple box in a UTM 31N projection space. The mesh had a minimum resolution of 1 km and resolution varied linearly with depth to a maximum of 1.5 km. The final mesh consisted of 1,035,877 nodes and 2,071,752 elements (Fig \ref{fig:doggerland_mesh}).

\subsection{Comparison to tsunami sediment deposits}

Given the uncertainties in the reconstruction of the palaeobathymetry, as well as in the drag coefficient and the Storegga tsunami itself, the simulations of the Shetland Isles and the Moray Firth were compared to the sedimentary record to assess the performance (Fig. \ref{fig:overviewmap}). To carry out this assessment a simple metric, which was the percentage of sites where sediment deposits have been recorded for the Storegga tsunami were inundated in the simulation, was calculated across both the Shetlands and Moray Firth simulations. It is also possible from the model to derive maximum flooding depth (i.e. run-up height) across the simulated domain to visually assess how close the wave came to inundating each sediment deposit under the different assumptions made (i.e. drag coefficient and palaeobathymetric reconstruction).

\section{Results}

Results across the Moray Firth and Shetland Isles local-scale models show that the inundation height of the Storegga tsunami as calculated by the models is generally lower than that derived from sediment deposits, even when using a very low drag coefficient, as many sites, especially in the Moray Firth do not get inundated in the model reconstructions. Seven sites out of 21 where a sediment deposit has been found show a tsunami inundation in the low drag simulations. Increasing drag decreased the height of the tsunami, as would be expected, meaning fewer sites were inundated (a decrease to just two sites). However, note that given the number of Storegga sediment deposits which are found within peat horizons, which indicates the presence of vegetation, the “medium” drag models are perhaps the most realistic. 

Increasing sea level by 5m produced more sites that show tsunami inundation as would be expected, but still not all sites are inundated (15 of 21). It is also worth noting that the increase of sea level by 5m does not lead to a uniform increase in peak wave height showing the non-linear nature of tsunami inundation. Decreasing sea level by 5m results in fewer sites being inundated, as expected (three sites from 21), with no inundation in the Moray Firth. 

\begin{sidewaystable}
\centering
    \caption{Maximum free surface height (MFS) and inundation coefficient ($I_c=0$ for no inundation, $1$ for inundated) for all nine modelled scenarios for each site of sedimentary deposit. Sites 1 to 10 are on the Shetland Isles, sites 11-21 are in the Moray Firth.}
\label{tab:results}
\begin{tabular}{l|l|ll|ll|ll|ll|ll|ll|ll|llll|}
    Site No. & Name & \multicolumn{2}{c|}{0\textbar L} & \multicolumn{2}{c|}{0\textbar M} & \multicolumn{2}{c|}{0\textbar H} & \multicolumn{2}{c|}{-5\textbar L} & \multicolumn{2}{c|}{-5\textbar M} & \multicolumn{2}{c|}{-5\textbar H} & \multicolumn{2}{c|}{+5\textbar L} & \multicolumn{2}{c|}{+5\textbar M}        & \multicolumn{2}{c|}{+5\textbar H} \\ \hline
        &  & MFS        & $I_c$         & MFS        & $I_c$         & MFS        & $I_c$          & MFS        & $I_c$          & MFS        & $I_c$          & MFS          & $I_c$         & MFS        & $I_c$          & MFS & $I_c$                        & MFS         & $I_c$          \\ \hline
    1 & Snarravoe, Unst & 14.80         & 1               & 8.18          & 1               & 0          & 0                & 9.11          & 1                & 1.99          & 1                & 0            & 0               & 18.16         & 1                & 13.15  & \multicolumn{1}{l|}{1}         & 0           & 0                \\
    2 & Burragarth, Unst & 8.80          & 1               & 6.31          & 1               & 0          & 0                & 0.61          & 1                & 0          & 0                & 0            & 0               & 16.56         & 1                & 14.25  & \multicolumn{1}{l|}{1}         & 0           & 0                \\
    3 & Norwick, Unst & 15.93         & 1               & 12.91         & 1               & 0          & 0                & 10.46         & 1                & 7.23          & 1                & 0            & 0               & 21.29         & 1                & 18.47  & \multicolumn{1}{l|}{1}         & 0           & 0                \\
    4 & Scasta Voe, Mainland & 0          & 0               & 0          & 0               & 0          & 0                & 0          & 0                & 0          & 0                & 0            & 0               & 1.33          & 1                & 0   & \multicolumn{1}{l|}{0}         & 0           & 0                \\
    5 & Garths Voe, Mainland & 0          & 0               & 0          & 0               & 0          & 0                & 0          & 0                & 0          & 0                & 0            & 0               & 0.43          & 1                & 0   & \multicolumn{1}{l|}{0}         & 0           & 0                \\
    6 & Otter Loch, Mainland & 0          & 0               & 0          & 0               & 0          & 0                & 0          & 0                & 0          & 0                & 0            & 0               & 0.51          & 1                & 0   & \multicolumn{1}{l|}{0}         & 0           & 0                \\
    7 & Sullum Voe, Mainland & 0          & 0               & 0          & 0               & 0          & 0                & 0          & 0                & 0          & 0                & 0            & 0               & 2.25          & 1                & 0   & \multicolumn{1}{l|}{0}         & 0           & 0                \\
    8 & Maggie Kettle's Loch, Mainland & 0          & 0               & 0          & 0               & 0          & 0                & 0          & 0                & 0          & 0                & 0            & 0               & 0          & 0                & 0   & \multicolumn{1}{l|}{0}         & 0           & 0                \\
    9 & Garth Loch, Mainland & 0.95          & 1               & 0          & 0               & 0          & 0                & 0          & 0                & 0          & 0                & 0            & 0               & 6.22          & 1                & 2.63   & \multicolumn{1}{l|}{1}         & 0           & 0                \\
    10 & Loch of Benston, Mainland & 0.70          & 1               & 0          & 0               & 0          & 0                & 0          & 0                & 0          & 0                & 0            & 0               & 5.79          & 1                & 2.64   & \multicolumn{1}{l|}{1}         & 0           & 0                \\ \hline
    11 & Wick River & 0          & 0               & 0          & 0               & 0          & 0                & 0          & 0                & 0          & 0                & 0            & 0               & 0          & 0                & 0   & \multicolumn{1}{l|}{0}         & 0           & 0                \\
    12 & Smithy House     & 0          & 0               & 0          & 0               & 0          & 0                & 0          & 0                & 0          & 0                & 0            & 0               & 0          & 0                & 0   & \multicolumn{1}{l|}{0}         & 0           & 0                \\
    13 & Creich     & 0          & 0               & 0          & 0               & 0          & 0                & 0          & 0                & 0          & 0                & 0            & 0               & 1.30          & 1                & 0.98   & \multicolumn{1}{l|}{1}         & 0.12           & 1                \\
    14 & Dounie     & 0          & 0               & 0          & 0               & 0          & 0                & 0          & 0                & 0          & 0                & 0            & 0               & 1.93          & 1                & 1.64   & \multicolumn{1}{l|}{1}         & 0.49           & 1                \\
    15 & Munlochy Bay     & 0.94          & 1               & 0.80          & 1               & 0.26          & 1                & 0          & 0                & 0          & 0                & 0            & 0               & 0.94          & 1                & 0.92   & \multicolumn{1}{l|}{1}         & 0.51           & 1                \\
    16 & Bellevue     & 0.68          & 1               & 0.54          & 1               & 0.09          & 1                & 0          & 0                & 0          & 0                & 0            & 0               & 1.06          & 1                & 0.94   & \multicolumn{1}{l|}{1}         & 0.30           & 1                \\
    17 & Tomich     & 0          & 0               & 0          & 0               & 0          & 0                & 0          & 0                & 0          & 0                & 0            & 0               & 1.09          & 1                & 0.97   & \multicolumn{1}{l|}{1}         & 0.31           & 1                \\
    18 & Barnyards     & 0          & 0               & 0          & 0               & 0          & 0                & 0          & 0                & 0          & 0                & 0            & 0               & 1.09          & 1                & 0.97   & \multicolumn{1}{l|}{1}         & 0.31           & 1                \\
    19 & Moniack     & 0          & 0               & 0          & 0               & 0          & 0                & 0          & 0                & 0          & 0                & 0            & 0               & 0          & 0                & 0   & \multicolumn{1}{l|}{0}         & 0           & 0                \\
    20 & Castle St., Inverness     & 0          & 0               & 0          & 0               & 0          & 0                & 0          & 0                & 0          & 0                & 0            & 0               & 0          & 0                & 0   & \multicolumn{1}{l|}{0}         & 0           & 0                \\
    21 &  Water of Philorth    & 0          & 0               & 0          & 0               & 0          & 0                & 0          & 0                & 0          & 0                & 0            & 0               & 0          & 0                & 0   & \multicolumn{1}{l|}{0}         & 0           & 0                \\ \hline
    $\dfrac{\sum I_c}{21}$     &     &               & 0.33       &               & 0.24       &               & 0.10        &               & 0.14        &               & 0.10        &                 & 0                  &               & \textbf{0.71}        &        & \multicolumn{1}{l|}{0.52} &                & 0.29       
\end{tabular}
\end{sidewaystable}

\subsection{Shetland Isles}

\begin{figure}[ht]
\centering
\includegraphics[width=0.5\textwidth]{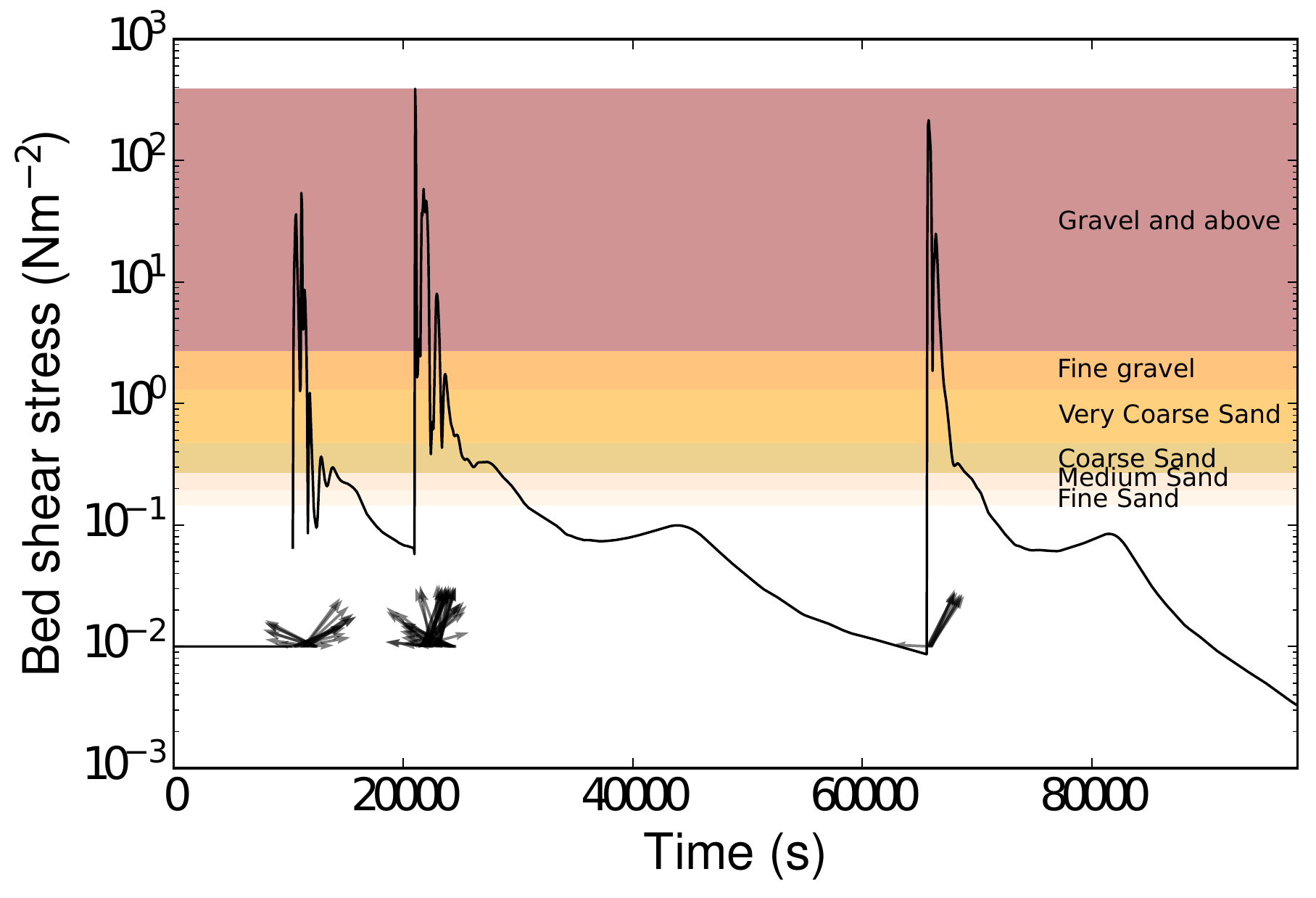}
    \caption{\label{fig:bss}Bed shear stress at Snarravoe (site 1). Coloured bands indicate critical bed shear stress for different grain sizes. Solid line indicates magnitude of bed shear stress and arrows indicate direction (with north aligned with the y-axis and east with the x-axis -- not scaled to magnitude).}
\end{figure}

The wave travels from the boundary, impacting the north-eastern coast of the Shetlands Isles (Unst) first, before travelling south, impacting Fetlar and Yell (see Video $1$ in supplementary information). The first wave travels inland eastward, inundating palaeovalleys around the site of the Norwick deposit. There is also inundation from the north along a palaeovalley and widespread inundation on low-lying land on the eastern side of Unst. On the southern side of Unst, a small wave first floods from the north (the original wave refracted around the northern margin of Unst), followed by a much larger wave from the south, which is the original wave refracted around the southern coast. These two waves meet around the location of the two deposits at Snarravoe and Burragarth, which appears to not quite be inundated by this wave. This original wave then travels south and a period of slow regression of the wave occurs. Before the first wave has fully receded, the second wave arrives, inundating the eastern side of Unst, but not as far inland as the original wave. However, this wave is much larger on the northern edge and the timing is such that the wave refracts around the southern and northern margins of Unst almost simultaneously, inundating the land at sites around Snarravoe and Burragarth. Burragarth appears to be inundated primarily from the north and Snarravoe from the south, though the flow dynamics at this time are complex (Fig. \ref{fig:bss}). This wave is also much larger on Yell and inundates the palaeovalley that cuts across Yell east to west (see Video $1$).

On the Mainland of Shetland, it is the fourth arrival that generates the greatest inundation distance around Sullum Voe (sites 5 to 8). The first to third wave arrivals all cause some inundation, but it is the timing of waves approaching the embayment from the north and east that causes the largest flooding extent. At sites 9 and 10, four inundations occur with the first and fourth causing the maximum flooding extent (Video $2$). 

\begin{figure}[ht]
\centering
\includegraphics[width=0.7\textwidth]{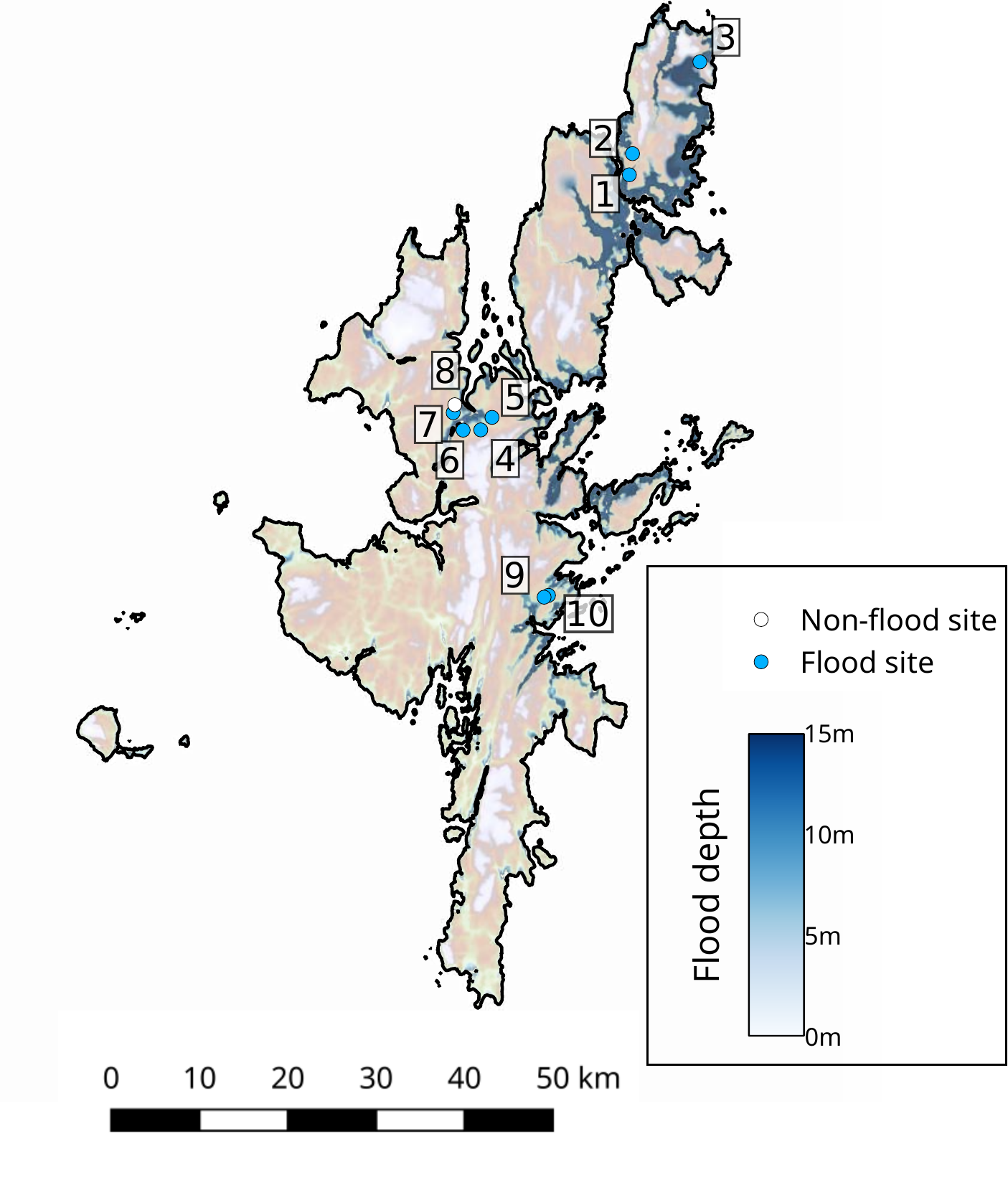}
\caption{\label{fig:shetlands_flooding}Maximum flood depth for the low-drag, +5 sea level simulation, in the Shetlands showing sites 1-10.}
\end{figure}

Many of the sites experience a high level of flooding: up to 16m in the base sea level case. If the sea level is increased by 5m, then an additional three sites are inundated and the maximum flow depth increases to over 22m (Fig. \ref{fig:shetlands_flooding}). Note this is an increase of over 6m, despite a 5m increase in sea level. Nearby sites experienced an increase in flow depth of less than 5m between the two scenarios. Increasing the drag results in lower flow depths as expected and an increase in drag from low to medium is roughly equivalent to a 5m drop in sea level. A 5m sea level drop results in only three sites being inundated (those on Unst).

The model here can also calculate bed shear stress. Here, a constant drag coefficient (0.0025) is applied to compute the bed shear stress in post-processing analysis. The results from site 1 (Snarravoe) show peak bed shear stress capable of moving boulders (as in Japan 2011), but this last for a very short time on the incoming wave (Fig. \ref{fig:bss}). The bed shear stress then rapidly drops on the outgoing wave and stays between 0.5 to 1 Nm$^{-2}$, which is around the critical bed shear stress for medium to fine sand. At sites 2 and 3 the results are analogous, and again there is a sustained period of bed shear stress higher than the critical shear stress for fine sand. This is consistent with the deposits found there \citep{Smith2004-xa}.

\subsection{Moray Firth}

\begin{figure}[ht]
\centering
\includegraphics[width=0.7\textwidth]{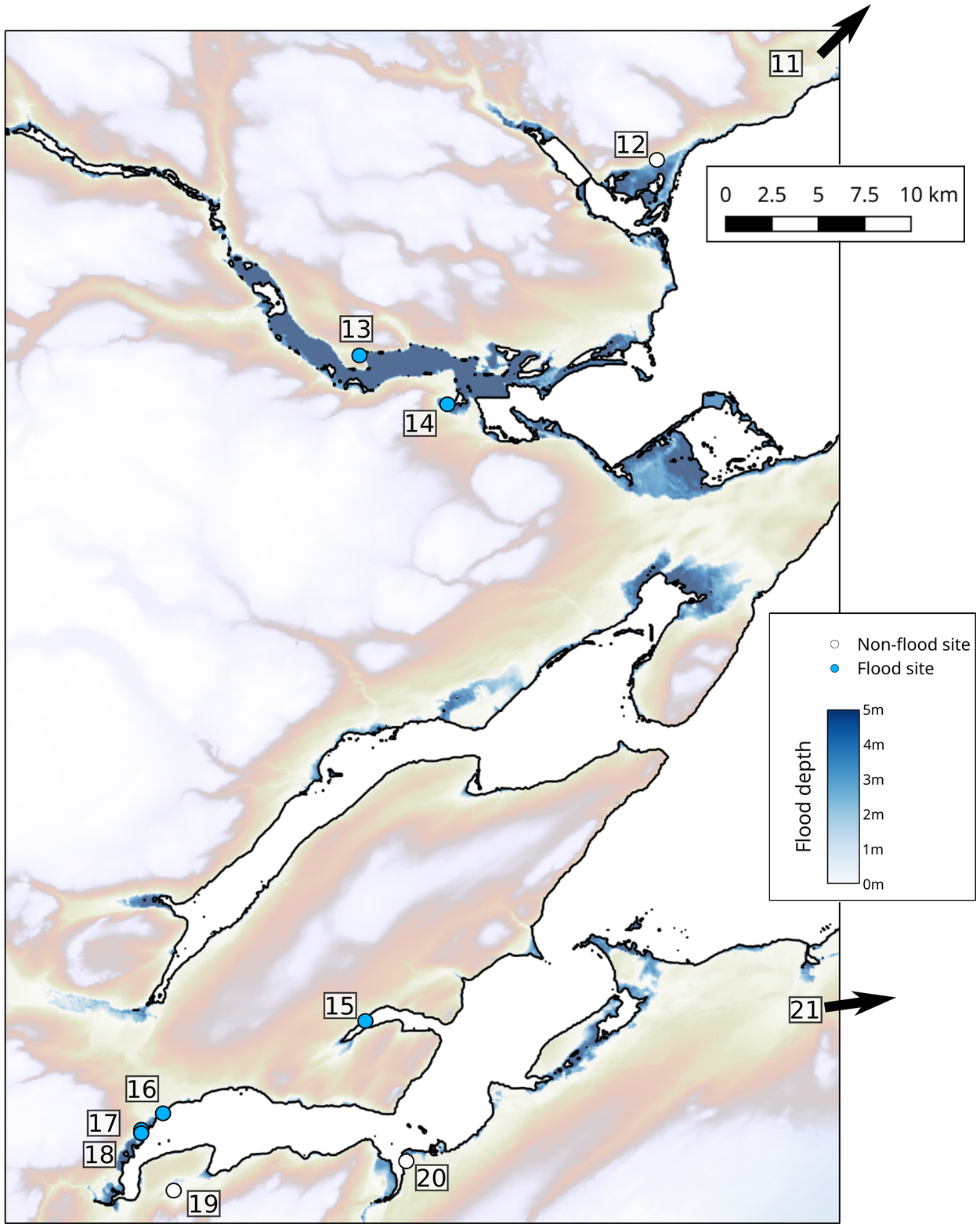}
\caption{\label{fig:moray_flooding}Maximum flood depth for the low drag, +5 sea level simulation, in the south-west corner of the Moray Firth showing sites 12-20. Sites 11 and 21 are off the map in the direction indicated by the arrows.}
\end{figure}

The wave history within the Moray Firth is somewhat simpler than around the Shetland Isles. The first wave is small and causes little inundation. There is then significant draw-down from the negative leading edge of the much larger second wave. The inundation occurs over a longer time than the wave that impacted the Shetlands and subsequently the receding phase is also longer, with some areas still inundated at the end of the simulation, particularly near Dornoch Firth (Video $3$, sites 13 and 14). A number of sites were not inundated in the Moray Firth simulations even with an additional sea level rise (Table \ref{tab:results}). However, two of these were near the northern and eastern boundaries of the domain (11 and 21, respectively). Most of the flooding at the sites of interest is small, certainly less than 1m flow depth (Fig \ref{fig:moray_flooding}).

As with the Shetlands, adding 5m to the reconstructed sea level means more sites are inundated (6 vs. 2), but again there is not an increase of 5m flow depth at sites that inundate in both scenarios and in fact the increase is less than 1m. In addition, many of the sites were close to flooding in the original simulation. A 5m drop in relative sea level shows no sites being flooded. Together these results show the non-linear nature of tsunami inundation predictions and the Moray Firth area requires either a larger wave or a slight increase in sea level in order to inundate the sedimentary deposit sites.

\subsection{Doggerland}

Based on the results from the sensitivity analysis from both the Moray Firth and Shetland simulations the low drag scenarios with either an additional 5m sea level increase or no additional sea level increase are the most plausible. The impact of the tsunami on Doggerland was therefore be assessed using these two scenarios.

\begin{figure}[ht]
\centering
\includegraphics[width=0.7\textwidth]{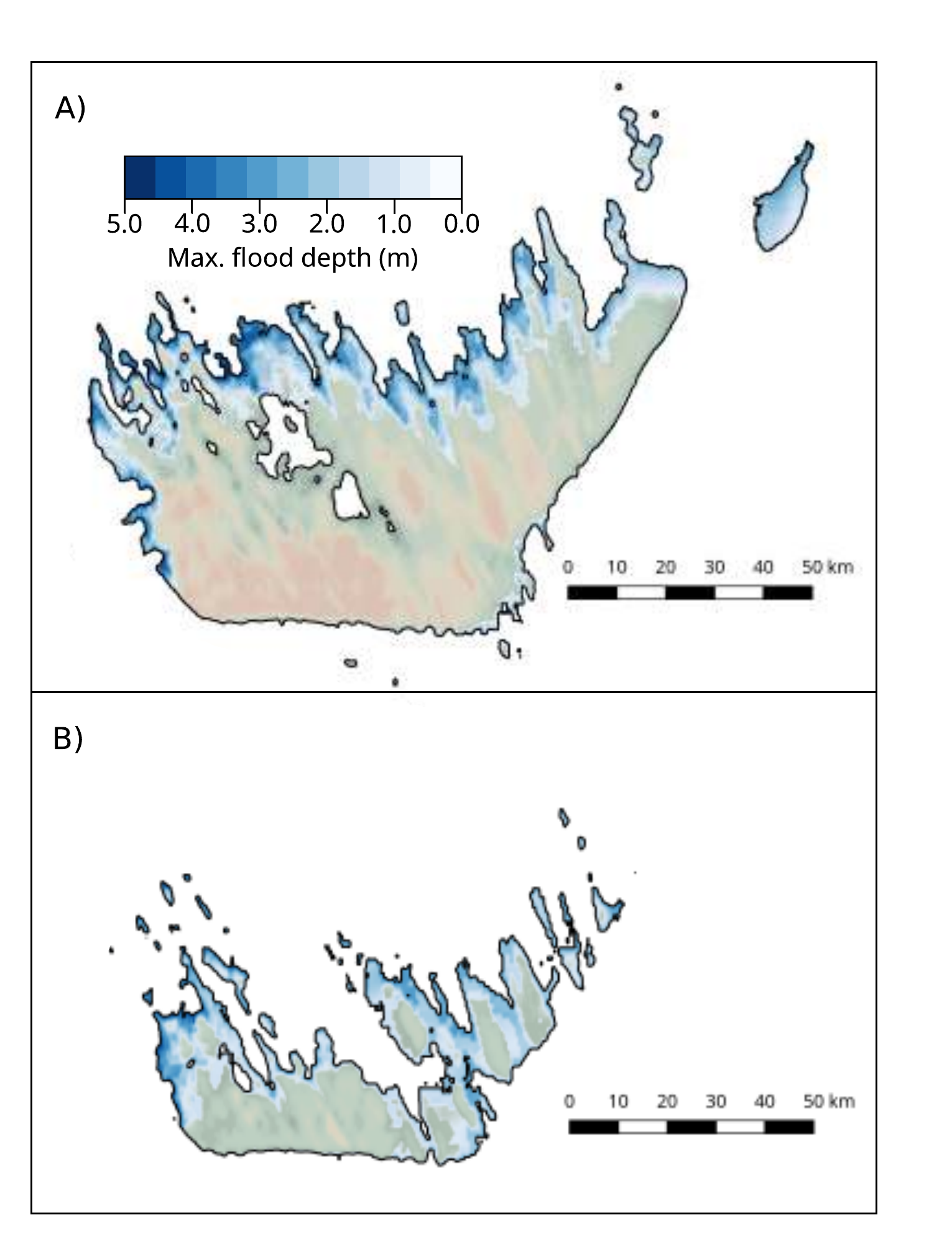}
    \caption{\label{fig:doggerland_flooding}Maximum flood depth on Doggerland using the zero sea level change (A) and +5m sea level scenario (B) and in both cases low drag. Green-brown colours show topography, blues show maximum flood depths.}
\end{figure}

The wave impacts the northern coast of the islands, completely inundating the smaller chain of islands recreated to the north east of the main island (Fig. \ref{fig:doggerland_flooding} and Video 4). The wave propagates up to 17 km inland in the north-east of the island and up to 21 km in the west of the island. Maximum flood depth is around 7 m and a total of 2000 km$^2$ was flooded (35 \% of the land surface). If relative sea level was 5m higher, then maximum flood depth increases to just over 9 m and the area flooded decreases to around 1200 km$^2$ (60 \% of the land surface). In both cases bed shear stress indicates that the tsunami would be capable of transporting cobble-sized grains (up to 256 mm). None of the scenarios modelled show a catastrophic inundation of Doggerland where the entire island was flooded. Doggerland would have to be substantially smaller or the Storegga wave substantially larger for the entire island to be flooded. Based on the topographic reconstructions here, the maximum runup height to entirely flood the island would be 16.5 m in the standard sea level scenario, indicating an increase of 7.5 m in the simulated wave is required.

\section{Discussion}

The work here demonstrates that a simple-solid block slide model of the Storegga tsunami can recreate the fundamental aspects of the event. Most sites in the Shetland Isles and the Moray Firth are inundated by this model, but this does require both a low drag coefficient and a change to the estimated relative sea level at the time. By simulating the inundation of the Storegga tsunami, insights into how the wave inundated the land surface can be gained. In the Shetlands, Unst in particular, two of the three sites flooded occur where the is confluence of waves approaching from the north and south simultaneously. This may provide ideal conditions for depositing tsunami sediments due to a sudden drop in flow velocities (Fig. \ref{fig:bss}). The model shows this by rapid decreases in bed shear stress. Similar sites from the model could indicate locations where more sedimentary deposits could be found. The locations where tsunami sediments are found but were not inundated by the model are generally in areas of rapidly changing coastlines, e.g. barriers and spit development, or where very close to inundation in one or more scenarios. Increased detail in reconstrucitons of the palaeocostaline would undoubtedly improve the simulations presented here.

The models also allow, for the first time, a recreation of the wave that inundated Doggerland. Whilst previous studies have focused on the catastrophic nature of the wave \citep{Weninger2008-zc,Hill2014-jo}. Here it is shown that the wave likely flooded only a small portion of the island. The modern tidal range at Lowestoft (the nearest modern tidal station) is 1.94m. The wave inundated to a level similar to this. Depending on the timing of the wave the Storegga tsunami may not have flooded much more than a spring high tide. A spring high tide of 2m would inundate nearly 1500 km$^2$ for the scenario considered with no additional sea level (25 \% of the land surface) and around 900 km $^2$ for the scenario with an additional 5m of sea level rise (45 \% of the land surface). In both cases the spring high tide is around 75\% of the tsunami inundation. Other natural hazards could produce similar if not larger inundation, for example the 2013 storm surge in the region caused wave heights of up to 6m above datum \citep{Spencer2014-cg}. Of course, this reconstruction is based on a simplistic reconstruction of Doggerland, which does not take into account changes in sedimentary regimes in the region in the last 8,000 years. Re-alignment of coastal regions during inundation show features such as barriers may be preserved, but in other cases these are transformed into thin veneers of beach sediments \citep{Forbes1995-xr}. A full study of the sedimentological and geophysical data at Doggerbank is required to provide a more realistic recreation of Doggerland.

The palaeobathymetric reconstruction is one of the main uncertainties in this study. Despite using a recent GIA model \citep{Bradley2011-kn}, more recent models show different histories of the British-Irish Ice Sheet and hence show moderate differences in sea level histories around the UK \citep{Kuchar2012-lz}. However, both GIA reconstructions don not replicate some recent sea level data from northern Scotland \citep{Long2016-nj}. To compensate for this uncertainty, a sensitivity analysis of adding or subtracting an additional 5m of sea level change shows that the Shetland Isles reconstructions were less sensitive to this than the Moray Firth reconstructions, particularly if relative sea level was 5m lower. As with Doggerland, tidal ranges are not included in these simulations. Modern tidal range at Lerwick (Shetlands) is 1.73m. Again, the wave would have to arrive co-incident with spring high tide in order to compensate partially toward the increased relative sea level change required to better match sedimentary records. There are no modern tidal gauge stations in the Moray Firth, but Wick on the north-east corner of the Scottish mainland has a similar tidal range to the Shetland Isles. However, it is not certain that a wave arrives at the Shetlands Isles co-incident with spring high tide and can then arrive in the Moray Firth valley co-incident with the spring high tide there. A more detailed study including astronomical tides is required. These tidal changes are based on modern tidal ranges. Tides around the UK have changed in the past 8,000 years and are predicted to do so with current sea level rise \citep{Pelling2014-vp}. Estimates for tidal ranges at the time of Storegga are not available in the regions of interest, but could alter by up to 0.5 m \citep{Shennan2000-aw}. A full reconstruction of tidal ranges at this time is required to reduce the uncertainty here. 

The method proposed here of comparing model output to sedimentary deposits by comparing if a site is inundated or not is perhaps the only viable method with current data. Comparing inundation between models and record with recent tsunami has been done using the observed inundation layer \citep{Lovholt2012-xz}. Similarly, for palaeotsunamis a comparison of the sand layer extent by either systematic coring or geophysical investigation would enable a much more thorough comparison between models and sedimentary data. It may also allow a recreation of the wave offshore using numerical inversion techniques \citep{Tang2015-uu}, which in turn could inform the initial tsunami source.

An alternative interpretation of these results is that the wave generated by the solid-block model is too small as it impacts the UK. Rather than requiring that the wave impacts the UK at spring high tide, a larger wave would inundate further inland and flood more of the sites where tsunami deposits are found. However, a larger wave may also change the size of the wave as it impacted the Norwegian coast. Here, there are also maximum wave heights as a series of lakes have recorded the tsunami inundation. A higher lake did not record any tsunami deposit, so in places a maximum inundation height can be obtained \citep{Romundset2011-qq}. A deformable slide model may generate a larger wave \citep{Smith2016-al} and may increase the performance of this reconstruction. This remains computationally challenging due to the high resolution required around the slide and to model the coastlines in sufficient detail.  

Similar studies on the Thera (Santorini) eruption and tsunami in the Aegean sea show a complex picture of a large-scale event potentially being the cause of the collapse of the Minoan civilisation \citep{Antonopoulos1992-pv}. Modelling of the event does show the potential of a large tsunami inundating several coastlines \citep{Novikova2011-xm}, though accurate dating of the sedimentary deposits may indicate the tsunami occurred a significant time before the collapse of the civilisation \citep{Minoura2000-qq}. Likewise, in this study, timing is again key; with several rapid pulses of sea level rise in the $\sim500$ years preceding the Storegga tsunami and the timing of Doggerland being abandoned. The modelling study here indicates that the ultimate cause of the flooding on Doggerland was likely sea level rise as the tsunami event was probably of comparable magnitude to storm events in the region. The pulses of sea level rise of decimetre to metre-scale in the 500 years prior to the Storegga event \citep{Lawrence2016-bm} likely inundated a substantial part of the island at the time and the tsunami wave was not large enough to be considered catastrophic. Studies similar to this and that of \citet{Novikova2011-xm} could clarify the potential causes of other ancient flooding myths across the globe.

\section{Conclusions}

A recreation of the Storegga tsunami, including both relative sea level changes and inundation has been carried out around the UK for the first time. The results show good agreement with sedimentary deposits found for the Storegga tsunami in the Shetlands Isles and the Moray Firth. The model gives indications of bed shear stress changes, which can provide insights into the flow dynamics that can result in the deposition of sediment. The model can also be used to recreate the tsunami on submerged landscapes. Here, the model indicates the tsunami impacted Doggerland with runup heights of up to 9 m, causing up to 2000 km$^2$ to be flooded (35 \% of the land surface). However, this flooding may be comparable to tides in that area at the time, and almost certainly comparable to large storm surges in the region. These results therefore show that whilst a major event in Doggerland, it is probable it was not a catastrophic event and sea level rise was the most likely cause of the abandonment of Doggerland.

\section{Acknowledgements}
The authors acknowledge support from NERC under grant NE/K000047/1. This work made use of the facilities of N8 HPC Centre of Excellence, provided and funded by the N8 consortium and EPSRC (Grant No. EP/K000225/1). The Centre is co-ordinated by the Universities of Leeds and Manchester. The authors thank Roland Gehrels for insighful and useful comments on an early draft of the manuscript.

\section{Supplementary Information}

Videos pertaining to this manuscript are deposited in figshare at DOI: \href{https://doi.org/10.6084/m9.figshare.5195035}{10.6084/m9.figshare.5195035}

\bibliographystyle{agsm}
\bibliography{biblio}

\end{document}